# A computational tool for trend analysis and forecast of the COVID-19 pandemic

Henrique Mohallem Paiva [1*], Rubens Junqueira Magalhães Afonso [2,3], Fabiana Mara Scarpelli de Lima Alvarenga Caldeira [1], and Ester de Andrade Velasquez [1]

[1] Institute of Science and Technology (ICT), Federal University of Sao Paulo (UNIFESP)
Rua Talim, 330, São José dos Campos, SP, Brazil

[2] Institute of Flight System Dynamics, Technical University of Munich (TUM)
München, Bayern, 85748, Germany

[3] Department of Electronic Engineering, Aeronautical Institute of Technology (ITA)
Praça Marechal Eduardo Gomes, 50, São José dos Campos, SP, Brazil

[*] Corresponding Author. E-mail: hmpaiva@unifesp.br

ORCID:
    https://orcid.org/0000-0001-7081-8383 (H. M. Paiva)
    https://orcid.org/0000-0001-9209-2253 (R. J. M. Afonso)

## Abstract

**Purpose:** This paper proposes a methodology and a computational tool to study the COVID-19 pandemic throughout the world and to perform a trend analysis to assess its local dynamics.

**Methods:** Mathematical functions are employed to describe the number of cases and demises in each region and to predict their final numbers, as well as the dates of maximum daily occurrences and the local stabilization date. The model parameters are calibrated using a computational methodology for numerical optimization. Trend analyses are run, allowing to assess the effects of public policies. Easy to interpret metrics over the quality of the fitted curves are provided. Country-wise data from the European Centre for Disease Prevention and Control (ECDC) concerning the daily number of cases and demises around the world are used, as well as detailed data from Johns Hopkins University and from the Brasil.io project describing individually the occurrences in United States counties and in Brazilian states and cities, respectively. U. S. and Brazil were chosen for a more detailed analysis because they are the current foci of the pandemic.

**Results:** Illustrative results for different countries, U. S. counties and Brazilian states and cities are presented and discussed.

**Conclusion:** The main contributions of this work lie in (i) a straightforward model of the curves to represent the data, which allows automation of the process without requiring interventions from experts; (ii) an innovative approach for trend analysis, whose results provide important information to support authorities in their decision-making process; and (iii) the developed computational tool, which is freely available and allows the user to quickly update the COVID-19 analyses and forecasts for any country, United States county or Brazilian state or city present in the periodic reports from the authorities.

## Keywords

## Highlights

- New mathematical model to describe and forecast the COVID-19 pandemic
- Innovative trend analysis approach
- Individual analysis for countries, U.S. counties and Brazilian states and cities
- Automatic calibrations, with no human intervention, and automatic data update
- Free computational tool available online

**Graphical Abstract**

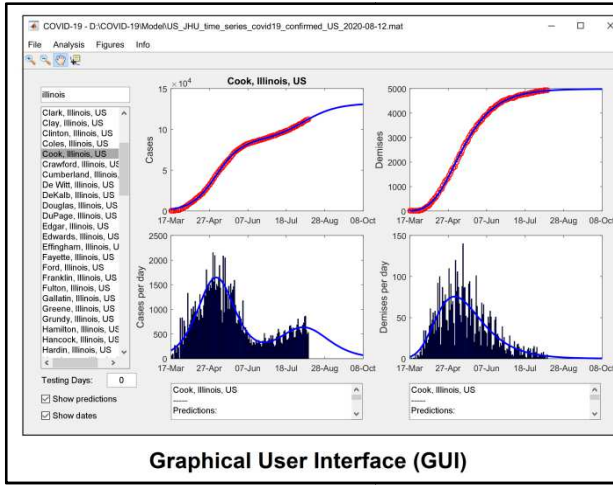

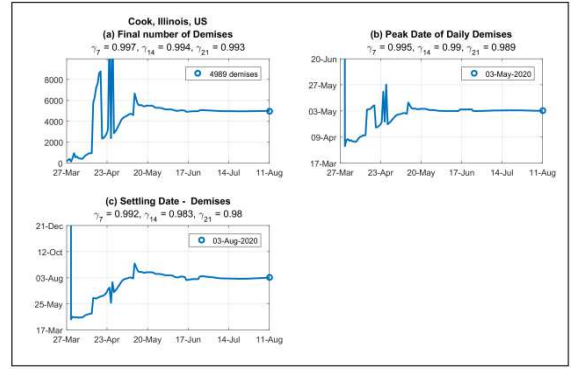

Trend Analysis

Graphical User Interface (GUI)

## 1. Introduction

On December 2019, a series of pneumonia cases of unknown cause emerged in Wuhan, China, with clinical presentations greatly resembling viral pneumonia [19]. The Chinese authorities identified a new type of coronavirus (novel coronavirus, named 2019-nCoV), which was isolated on 7 January 2020 [51]. Coronaviruses are a family of viruses that can cause respiratory, hepatic, and neurological diseases in humans and animals [13].

Initially, these infections were thought to result from zoonotic (animal-to-human) transmission. However, an exponential growth of case incidence, with many cases detected in other parts of the world, showed a strong evidence of human-to-human secondary transmission [27]. On March 2020, the coronavirus disease (COVID-19) was declared a public health emergency of international concern by The World Health Organization [49]. A worrying aspect in relation to the disease is the fact that it is highly contagious, spreading very quickly and causing overcrowding in the health system [52].

The main symptoms reported are fever, dry cough, and sore throat, with cases of severe pneumonia. These clinical events are associated with admission to the intensive care unit and have a high mortality rate [19][51].

The nucleic acid test or genetic sequence for 2019-nCoV is considered the standard method to confirm the infection. A laboratory-confirmed case is defined as a positive result, using high-throughput sequencing or real-time reverse transcriptase-polymerase chain reaction (RT-PCR) with nasal and pharyngeal swab specimens [15]. It is important to note that the virus can sometimes be detected in blood tests and in rectal swabs, but not found in the throat swab (false negatives), with the result that these patients can act as carriers and transmit the infection [37].

Since January 2020, many studies have been done to assess the transmission potential of 2019-nCoV nationally and internationally, as well as to forecast its spread. Quarantine measures have been implemented worldwide, as international travel has helped to spread the virus to other parts of the world [7][52]. At the time of this writing, more than 20 million people worldwide were infected with COVID-19 and more than 740,000 died. Many countries have community transmission, defined as a major outbreak of local transmission [50].

Many efforts have been done to determine an effective treatment and to develop a vaccine [6][16][30]. It is expected that infection with 2019-nCoV will result in some form of immunity, but there is not enough evidence that the antibody-mediated immunity will guarantee protection against a second infection [39]. Currently, the results of many studies suggest that early detection, hand washing, self-isolation and household quarantine are effective to mitigate this pandemic [7].

Similar situations, although in smaller scale, occurred during the Influenza A epidemic in 2009 and during the Middle East Respiratory Syndrome coronavirus (MERS-CoV) epidemic in 2012. The Influenza A virus appeared in April 2009 [45] and caused a pandemic with more than 280,000 deaths worldwide [43]. The MERS epidemic emerged in Saudi Arabia in 2012 [22] and caused thousands of infections in dozens of countries worldwide. This virus, also belonging to the coronavirus family, has a high fatality rate and its symptoms include severe acute pneumonia [5]. Under these scenarios, Dugas *et al*. [11] created a forecasting model for Influenza A and Kim *et al*. [22] formulated a forecasting model for MERS transmission dynamics and estimated transmission rates, considering several categories of patients and transmission rates. In fact, mathematical models have been widely used to study the transmission dynamics of infectious diseases, enabling the understanding of the disease spread and the optimization of disease control [34].

Forecasting models are used to predict future behavior as a function of past data. This is a widely used method in the implementation of epidemic mathematical models, since it is necessary to know the past behavior of a disease to understand how it will evolve in the future. Accurate forecasts of disease activity could allow for better preparation, such as public health surveillance, development and use of medical countermeasures, and hospital resource management [8].

A similar approach is the concept of trend analysis, which allows predicting future behavior with accuracy, especially in the short run. A trend is a change over time exhibited by a random variable [28]; trend analyses provide direction to a trend from past behavior, allowing predicting future data. For better effectiveness, the predictions should be updated periodically, as soon as new data are available.

The technique of trend analysis is widely used in several areas of science, such as finances [2][48] and meteorology [28] [36]. In the context of health systems, trend analysis was used by Zhao *et al.* [57], to analyze malignant mesotheliomas in China, aiming to provide data for its prevention and control; by Soares *et al.* [46], to predict the testicular cancer mortality in Brazil; by Zahmatkesh *et al.* [56], to forecast the occurrences of breast cancer in Iran; by Mousavizadeh *et al.* [33], to forecast multiple sclerosis in a region of Iran; and by Yuan *et al.* [55], to analyze and predict the cases of type 2 diabetes in East Asia.

Modeling and prediction of the dynamics of the COVID-19 pandemic is a subject of great interest. Therefore, a myriad of papers on this theme have been published over the last months, exploiting different modeling approaches, such as compartment models [18], time series analysis [44], artificial intelligence [41][54], and regression-based models [17][40]. For this purpose, some research groups extended previous epidemiological models to describe the COVID-19 pandemics: Lin *et al.* [25] created a conceptual model for the COVID-19 outbreak in Wuhan, China, using components from the 1918 influenza pandemic in London, while Paiva *et al.* [38] proposed a dynamic model to describe the COVID-19 pandemic, based on a model previously developed for the MERS epidemic. This list is far from being exhaustive. For a detailed survey on different modeling approaches in this context, the reader is referred to review papers such as [26] and [32].

It is important to note that the behavior of the pandemic may vary greatly in the different regions of the world, due to characteristics such as different social habits (higher or lower physical interaction between citizens), capacity of the local health system, different governmental actions, and so on. Therefore, the parameters of a mathematical model need to be tailored to the region where the disease behavior is being studied. Furthermore, even in the same region, the conditions may vary very quickly, in a matter of weeks or even days (for instance, following the decree or release of a lockdown, or the saturation of the available intensive care unit vacancies in the hospitals); thus, the model parameters would need to be updated very often, usually by an expert. However, these analyses might take time and require dedicated work from highly qualified personnel, thus decreasing their availability. It is natural to expect that such analyses are run periodically at the country level, but the same may not be a reality locally at every municipality. Therefore, in this scenario, it is useful to have a computational tool to perform a quick and automatic analysis and forecast of the disease conditions in any region, following the periodic updates published by the authorities. This is the purpose of the present paper.

## 2. Methods

### 2.1. *Mathematical formulation of the curve to describe the data*

In the present paper, the fundamental curve that is used to describe the historical data is an asymmetric sigmoid, i.e., letting the independent variable be t, then the dependent variable is given as a function $f: \mathbb{R} \mapsto \mathbb{R}^+$ [42]:

$$f(t) = \frac{A}{\left(1 + \nu e^{-\frac{(t-t_p)}{\delta}}\right)^{\frac{1}{\nu}}} \quad (\,1\,)$$

with the parameters $A \in \mathbb{R}^+ \cup \{0\}$, $\nu \in \mathbb{R}^+$, $\delta \in \mathbb{R}^+$, $t_p \in \mathbb{R}^+ \cup \{0\}$. In the present work, the independent variable t is the time in days, whereas the dependent variable is either the cumulative number of individuals that were positively tested for SARS-CoV-2 or the cumulative number of individuals deceased with the disease as the cause. Notice that

$$\lim_{t \to \infty} f(t) = A \quad (\,2\,)$$

i.e., the modeling of the cumulative number of cases/demises by ( 1 ) implies convergence to a final value A. However, the convergence is asymptotic, therefore it is interesting to know when a certain threshold of the final number of infected/deceased has been reached. For that purpose, let a time instant $\tau_\alpha$ be such that a particular value $f(\tau_\alpha)$ is reached:

$$f(\tau_\alpha) = \alpha A \quad (\,3\,)$$

where the parameter $\alpha \in \,]0,1[$. Then, by replacing ( 1 ) for $f(\tau_\alpha)$ in ( 3 ), one may solve to find:

$$\tau_\alpha = t_p - \delta \ln\left\{\frac{1}{\nu}\left[\left(\frac{1}{\alpha}\right)^\nu - 1\right]\right\} \quad (\,4\,)$$

Therefore, from ( 4 ) one can determine the (finite) instant when a certain proportion of the final number of cases/demises is reached, which is a useful figure to evaluate whether the contamination can be considered over or not. In this paper, the settling date of the contamination is adopted as $\tau_{0.98}$, corresponding to the day where the number of occurrences reaches 98% of its final value. The settling ratio of 98% is a standard value used in the analysis of dynamic systems [10].

The rate at which the number of infections/demises grows can be calculated by differentiation of ( 1 ) with respect to the independent variable t, which yields

$$\frac{df(t)}{dt} = \frac{A}{\delta} \frac{e^{-\frac{(t-t_p)}{\delta}}}{\left(1 + \nu e^{-\frac{(t-t_p)}{\delta}}\right)^{\frac{\nu+1}{\nu}}} \quad (\,5\,)$$

As a matter of fact, the value of ( 5 ) in a particular day $t$ is an important indicator for healthcare infrastructure decision-making concerning the number of infected individuals, as a higher value indicates that the upcoming period might stress the healthcare infrastructure, whereas a comparatively lower value points that the number of new cases might be accommodated with the existing infrastructure. By analyzing the number of individuals that are cured each day and discharged from the facilities and comparing it with the rate of newly infected individuals, if the first is greater than the latter, than the capacity of the facilities is enough to treat the ill and they will not be endangered by lack of proper treatment.

Differentiating ( 5 ) with respect to $t$ yields

$$\frac{d^2f(t)}{dt^2} = \frac{Av}{\delta^2} \frac{e^{-\frac{(t-t_p)}{\delta}}\left(1 - e^{-\frac{(t-t_p)}{\delta}}\right)}{\left(1 + ve^{-\frac{(t-t_p)}{\delta}}\right)^{\frac{2v+1}{v}}} \qquad (\ 6\ )$$

A sign change in ( 6 ) occurs when the term $\left(1 - e^{-\frac{(t-t_p)}{\delta}}\right)$ crosses zero, as the remaining terms are all positive for any $t \in \mathbb{R}$. Therefore, there is a single inflection in the curve ( 5 ) at $t = t_p$. On the other hand, since $\frac{d^2f(t)}{dt^2} > 0$ for $t < t_p$ and $\frac{d^2f(t)}{dt^2} < 0$ for $t > t_p$, this point corresponds to the maximum rate, i.e., the daily number of either infected or deceased individuals. Replacing $t = t_p$ in ( 1 ) yields

$$f(t_p) = \frac{A}{(1+v)^{\frac{1}{v}}} \qquad (\ 7\ )$$

Notice from ( 7 ) that $v = 1$ entails $f(t_p) = 0.5A$, i.e., the sigmoid curve crosses half of the final value at $t = t_p$. This is deemed a symmetric sigmoid. For the sake of understanding, consider two other illustrative possible values of $v$:

a) for $v = 2$, from ( 7 ), $f(t_p) = A/\sqrt{3} \approx 0.58A$, that is, the inflection happens at a later stage, when roughly 58% of the final values has been reached;

b) for $v = 0.5$, from ( 7 ), $f(t_p) = 4A/9 \approx 0.44A$, in other words, the inflection happens at an earlier stage, when approximately only 44% of the final values has been reached.

It is clear from these examples and from ( 1 ) that the value of $v$ controls the degree of asymmetry in the sigmoid curve, with $v = 1$ representing a symmetric curve about the $t = t_p$ vertical straight-line. This is illustrated in Figure 1(a), where ( 1 ) is shown for three values of $v$ whereas Figure 1(b) shows ( 5 ), i.e., the rate. It is interesting to remark that the value of $v$ impacts the symmetry of the derivative, with $v = 1$ representing a Gaussian curve, with acceleration and deceleration phases occurring at the same rate. When $v < 1$, the deceleration phase of the sigmoid is slower than the acceleration phase; when $v > 1$, the opposite occurs.

For their capability of representing processes with asymmetric acceleration and deceleration phases, asymmetric sigmoid curves are interesting to represent the data of a pandemic. Many factors can contribute to the asymmetry between acceleration and deceleration phases besides the very nature of the disease spread, such as the introduction of policies by health authorities in order to slow down the spread, e.g., reduced social contact. Therefore, this extra degree of freedom brought by the asymmetric sigmoid curve is useful to better represent the data. Moreover, the added complexity with regard to a symmetric curve is due only to the necessity of estimating a single additional parameter, namely $\nu$.

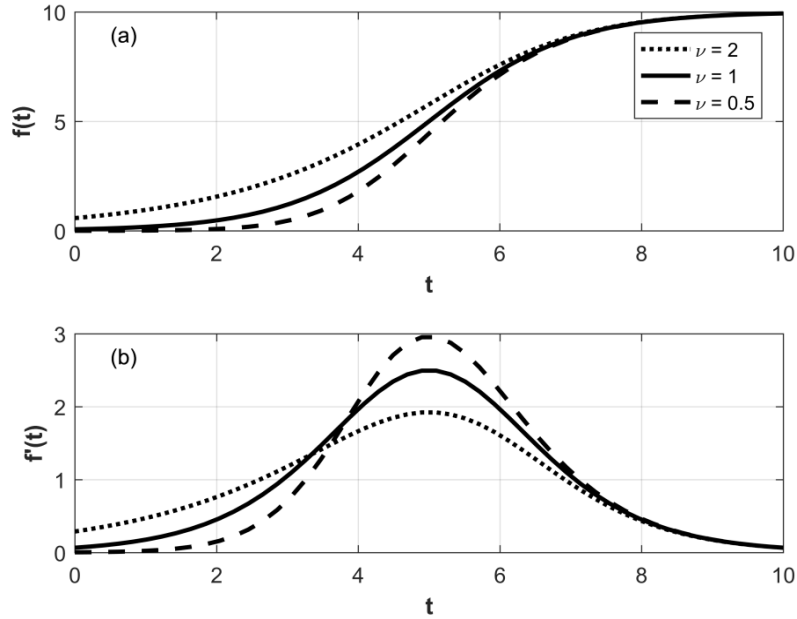

**Figure 1 – (a) Sigmoid curves and (b) their derivatives for different values of the parameter $\nu$. The remaining parameters are $A = 10$, $t_p = 5$, and $\delta = 1$.**

In our context, there are three main sigmoid parameters of interest, which are described in Table 1.

**Table 1 – Main sigmoid parameters of interest**

| Parameter | Eqs. | Description |
|-----------|------|-------------|
| A | ( 1 )( 2 ) | Final number of occurrences. |
| $t_p$ | ( 1 ) | Date when the maximum number of daily occurrences is achieved. |
| $\tau_{0.98}$ | ( 3 )( 4 ) | Settling date, defined as the date when the number of occurrences reaches 98% of its final value. |

The next section presents the algorithm used to estimate the parameters $A$, $\nu$, $\delta$, and $t_p$ based on measured data from either the number of newly infected individuals per day or the number of deceased per day.

### *2.2. Parameter estimation*

The parameters A, ν, δ, and $t_p$ are estimated based on the solution of a constrained optimization problem, in which the Integral Time Square Error (ITSE) [10] is minimized, where the error is the difference between the value of f(t) output by ( 1 ) and the corresponding data y(t) obtained from the authorities at the same day. We consider a time window for $t \in \{0, 1, \ldots, t_{end}\}$ for which the data y(t) are available at each day. There is a small abuse of notation by restricting the real-valued variable t to assume only integer values coinciding with the number of the day,

Let the vector of parameters to be estimated be defined as

$$\Theta = [A \quad ν \quad δ \quad t_p]^T \qquad ( 8 )$$

where the symbol $•^T$ indicates the transpose of a vector •. The optimal value of the vector Θ is given as

$$\Theta^* = \underset{\{A \geq 0, ν > 0, δ > 0, t_p \geq 0\}}{\text{argmin}} \sum_{t=0}^{t_{end}} t[y(t) - f(t, \Theta)]^2 \qquad ( 9 )$$

where the argument Θ was explicitly included in f(t, Θ) to emphasize that the parameters may be varied during the optimization process. Note that, for optimization purposes, strict inequalities cannot be implemented, therefore for the constraints ν > 0 and δ > 0, an arbitrary small positive real number μ > 0 is chosen and the constraints are approximated as ν ≥ μ and δ ≥ μ. After the optimization problem is solved to yield Θ*, the optimal values of $A^*, ν^*, δ^*, t_p^*$ are fixed values used to build the curve.

The function f(t, Θ) is nonlinear in the parameters Θ, and the cost function exacerbates that further, rendering the optimization problem nonlinear. Moreover, the inequality constraints introduce additional difficulty, rendering the analytical solution of the optimization problem impractical. Therefore, numerical methods must be used.

One class of methods that are suitable for nonlinear constrained optimization is the so-called Sequential Quadratic Programming (SQP) [14][20][21]. SQP iteratively approximates the general nonlinear cost function in ( 9 ) by a quadratic one, and the constraints by linear ones, which entails a Quadratic Programming (QP) problem. QPs can be solved to global optimality in finite time, therefore each iteration of the SQP method takes finite time. The solution of the underlying QP approximation is then used to build a next iterate, for which another QP is solved, therefore the name Sequential Quadratic Programming. SQP presents good convergence properties, converging quadratically to the optimal solution when the active set does not change [35]. The implementation of SQP that is used in the present work is that of the function **fmincon** [31], from the Optimization Toolbox$^{TM}$ of MATLAB®.

### *2.3. Multiple sigmoids*

A second wave of spread has not been discarded. On the contrary, researchers argue that lifting the social distance measures might indeed lead to a retake in the infections [1][24][29][53].

In order to describe the occurrence of multiple epidemiological waves, we propose to employ a sum of sigmoids. For this purpose, let $N_s$ be the adopted number of sigmoids. Equation ( 1 ) is then generalized to

$$f(t) = \sum_{i=1}^{N_s} f_i(t) \quad ( 10 )$$

where

$$f_i(t) = \frac{A_i}{\left(1 + v_i e^{-\frac{(t-t_{p,i})}{\delta_i}}\right)^{\frac{1}{v_i}}}, \quad i = 1, 2, \ldots, N_s \quad ( 11 )$$

Similarly, the vector of parameters $\Theta$, originally given by ( 8 ), is generalized to a column vector with $4N_s$ parameters defined as:

$$\Theta = \begin{bmatrix} \Theta_1^T & \Theta_2^T & \ldots & \Theta_{N_s}^T \end{bmatrix}^T \quad ( 12 )$$

where

$$\Theta_i = [A_i \quad v_i \quad \delta_i \quad t_{p,i}]^T, \quad i = 1, 2, \ldots, N_s \quad ( 13 )$$

With these extended definitions, equation ( 9 ) can still be used to estimate the value of $\Theta^*$ by considering the inequalities applied to each $A_i$, $v_i$, $\delta_i$ and $t_{p,i}$, $i = 1, 2, \ldots, N_s$.

It is important to establish the number of sigmoids $N_s$. For this purpose, an evaluation of the number of switches between acceleration and deceleration phases is performed. The rationale behind this assessment is: each sigmoid results in a single acceleration and a single deceleration phases, with a clear switching point between them, as discussed in Section 2.1. Therefore, the number of sigmoids can be estimated by counting the amount of such switches between acceleration and deceleration phases. However, this counting requires careful consideration, as one is dealing with real noisy data. More so, recall that for identifying acceleration/deceleration the second derivative of the cumulative number of either infected or deceased individuals has to be considered. As it is well known, differentiation is prone to increase the effect of noise in the measurements [10]. Therefore, to mitigate the effect of noise in increasing artificially the amount of switches, a common approach is to consider a deadzone [10] in the difference between the acceleration and deceleration. Let S be the set of switching instants between acceleration and deceleration phases, then, for each $t = 0, 1, \ldots, t_{end} - 1$, the following logic is used to implement an identification of switches with a deadzone:

$$\frac{d^2f}{dt^2}(t) \geq \epsilon \text{ and } \frac{d^2f}{dt^2}(t+1) \leq -\epsilon \Rightarrow t \in S$$
$$\frac{d^2f}{dt^2}(t) \leq -\epsilon \text{ and } \frac{d^2f}{dt^2}(t+1) \geq \epsilon \Rightarrow t \in S \quad ( 14 )$$
$$\text{otherwise}, t \notin S$$

where the parameter $\epsilon$ can be adjusted to provide a compromise between noise and detection sensitivity. In the present work, the value was set to $\epsilon = 3 \cdot 10^{-5}$ persons/day$^2$.

Thus, the number of switches is given by the cardinality of S

$$N_s = |S| \quad (15)$$

Recall, from Table 1, that there are three parameters of interest. The final number of occurrences may be obtained as:

$$A = \sum_{i=1}^{N_s} A_i \quad (16)$$

On the other hand, when a sum of sigmoids is used, there are no analytical expressions to determine the other two parameters of interest, i.e., the date of maximum number of daily occurrences $t_p$ and the settling date $\tau_{0.98}$. In this case, a numerical search algorithm has to be used to find each of these parameters.

The optimization problems to determine these parameters can be posed as follows:

$$t_p = \frac{\text{argmax}}{t \geq 0} \frac{df}{dt}(t) \quad (17)$$

$$\tau_{0.98} = \frac{\text{argmin}}{t \geq 0} [f(t) - 0.98A]^2 \quad (18)$$

These two optimization problems are solved using the Nelder-Mead algorithm [23]. It should be noted that each problem has only one independent variable (time). Therefore, the search algorithm converges very quickly to the desired solution.

The rationale to select the use of one or multiple sigmoids will be explained in a following section, which discusses the complexity of the model.

### 2.4. Criteria for statistical analysis of the matching between the fitted curve and the data

Two criteria are used to evaluate the degree of fidelity of the fitted curves to the data. The first is the so-called Root Mean Square Error (RMSE), defined as:

$$RMSE = \sqrt{\sum_{t=0}^{t_{end}} \frac{[y(t) - f(t, \theta^*)]^2}{t_{end} + 1}} \quad (19)$$

From ( 19 ) the name of RMSE becomes clear, as it involves the square root of the mean of the squared error. Notice that, in ( 19 ), the values of the curve with the optimal parameters $f(t, \theta^*)$ are used to calculate the error between the data and the value returned by the fitted curve. Moreover, the term $t_{end} + 1$ reflects the number of terms in the summation, as the index $t$ starts at $0$ and ends at $t_{end}$. The RMSE is used in statistical analysis to measure compactly the degree of fidelity between the fitted curve and the data. The lower the value of the RMSE, the better the fitted curve matches the data [3].

In this paper, a normalized version of the RMSE is used, obtained as:

$$\text{normalized RMSE} = \frac{RMSE}{A} \quad (20)$$

where A is the final number of occurrences, as defined in ( 1 ) and ( 16 ) for one and multiple sigmoids, respectively. This normalization is adopted to allow a fair comparison of the RMSE of different curves.

A second criterion to determine the quality of the representation of the data by the fitted curve generally applied in statistics is the squared correlation coefficient, which varies between 0 and 1, with the latter meaning that there exists a perfect linear functional relationship between the data and the fitted curve points, whereas the first means the opposite. First, let us define the covariance of the data as

$$cov[y(\bullet), f(\bullet, \Theta^*)] = \sum_{t=0}^{t_{end}} \frac{[y(t) - \mu_y][f(t, \Theta^*) - \mu_f]}{t_{end}} \quad (\ 21\ )$$

where $\mu_y$ and $\mu_f$ are the mean values of $y(t)$ and $f(t, \Theta^*)$, respectively, i.e.

$$\mu_{\bullet} = \sum_{t=0}^{t_{end}} \frac{\bullet}{t_{end} + 1} \quad (\ 22\ )$$

in which the symbol $\bullet$ can be replaced by either of $y(t)$ and $f(t, \Theta^*)$, yielding $\mu_y$ and $\mu_f$, respectively. Similarly, the variances of $y(t)$ and $f(t, \Theta^*)$ are

$$var[y(\bullet)] = cov[y(\bullet), y(\bullet)] = \sum_{t=0}^{t_{end}} \frac{[y(t) - \mu_y]^2}{t_{end}} \quad (\ 23\ )$$

$$var[f(\bullet, \Theta^*)] = cov[f(\bullet, \Theta^*), f(\bullet, \Theta^*)] = \sum_{t=0}^{t_{end}} \frac{[f(t, \Theta^*) - \mu_f]^2}{t_{end}} \quad (\ 24\ )$$

The squared correlation coefficient $R^2$ can then be determined from ( 21 )-( 24 ) as:

$$R^2 = \frac{\{cov[y(\bullet), f(\bullet, \Theta^*)]\}^2}{var[y(\bullet)]var[f(\bullet, \Theta^*)]} \quad (\ 25\ )$$

### 2.5. Criteria for assessment of convergence of the sigmoid towards the final value

Additional criteria are defined to evaluate whether the data are enough to allow the convergence of the estimated values of the parameters $\Theta^*$. This is carried out by fitting the sigmoid curves to the data for each possible value of $n \in \{10, 11, \ldots, t_{end} - 21, t_{end} - 20, \ldots, t_{end}\}$. Thus, instead of using all available data as in ( 9 ), windows of varying length are used; the minimum length of a window is adopted as 10 to ensure a minimum amount of data to calibrate the curve. Therefore, the sigmoid parameters $\Theta$ are estimated within different windows as

$$\Theta^*(n) = \begin{array}{c} \text{argmin} \\ \{A_i \geq 0, \nu_i > 0, \delta_i > 0, t_{p,i} \geq 0\} \end{array} \sum_{t=0}^{n} t[y(t) - f(t, \Theta)]^2 \quad (\ 26\ )$$

where $i = 1, 2, \ldots, N_s$, depending on the number of sigmoids.

The main parameters in Table 1 are then determined from $\Theta^*(n)$ as follows:

- For a single sigmoid, $A^*(n)$ and $t_p^*(n)$ are directly extracted from $\Theta^*(n)$ in view of ( 8 ), whereas $\tau_\alpha^*(n)$ is calculated by ( 4 ) employing $t_p^*(n)$, $\delta^*(n)$ and $\nu^*(n)$ extracted from $\Theta^*(n)$ considering ( 8 ).

- For multiple sigmoids, ( 16 )-( 18 ) are used to determine $A^*(n)$, $t_p^*(n)$ and $\tau_{0.98}^*(n)$.

Then, the relative variation of the estimated values of these parameters is calculated for each time window and multiplied over the time window, composing indices to evaluate if the data are enough to asseverate the suitability of the sigmoid that was fitted. These indices are defined as

$$\gamma_k^\bullet = \prod_{j=1}^{k} \min \left( \frac{\bullet^* (t_{end} - j - 1)}{\bullet^* (t_{end} - j)}, \frac{\bullet^* (t_{end} - j)}{\bullet^* (t_{end} - j - 1)} \right) \quad (\ 27\ )$$

where the symbol $\bullet$ represents one of the parameters of interest, namely, $A^*(n)$, $t_p^*(n)$, or $\tau_\alpha^*(n)$, for a time window up to k days of data. It is clear that, if the data are enough and a suitable set of parameters is found, then each of the terms in the product in ( 26 ) approaches one. Therefore, the closer the value $\gamma_k^\bullet$ is to one, the better the fit. Moreover, the "min" in ( 26 ) ensures that each term in the product is less than or equal to one, from which it follows that $\gamma_k^\bullet \leq 1$. Analyzing $\gamma_k^\bullet$ for different values of k enables the conclusion of whether the convergence has occurred or not within variable window sizes. We adopt windows of size 7, 14 and 21 days, in order to verify the stability of the predictions over the last one, two and three weeks.

### 2.6. Selecting the complexity of the model

From the previous discussion, it is possible to choose among different curve types (symmetric or asymmetric) and numbers (single or multiple sigmoids). This plays an important role both in the accuracy of the fit and in the complexity of the models (as per the different amounts of parameters to be estimated with each choice).

It should be noted that the choice of a more complex model without a significant increase in the accuracy may lead to the problem of model overfitting, that is, an exaggeration while fitting of the training data that may compromise the generalization of the model predictions [47]. In order to avoid this problem, criteria should be established to enable a compromise between accuracy and complexity. These criteria are described in this section.

### 2.6.1 Symmetric or assymetric curve

Particularly for cases of regions where the contagion is in its early stage, there are not enough data to observe a deceleration phase. Therefore, in this case the data are insufficient to support estimation of the asymmetric curves. In these situations, the symmetric curves can be used in the fitting and an automated decision of whether to present results with a symmetric or an asymmetric curve has to be done. The criterion for this decision considers a compromise between complexity and quality of the fitting results. The complexity is deemed higher for the asymmetric curve, as it requires estimation of one additional parameter, namely $v$. As for the quality, it is evaluated through the following ratio:

$$\eta_1 = \frac{1 - R_a^2}{1 - R_s^2} \qquad (\ 28\ )$$

where $R_a^2$ represents the squared correlation coefficient obtained from fitting an asymmetric curve and $R_s^2$ the one yielded by a symmetric curve.

Recall that the value of the squared correlation coefficient ranges from zero to one and that this latter value implies a perfect relationship between the data and the curve. Furthermore, the symetric sigmoid may be considered a particular case of the asymetric sigmoid, obtained by imposing $v = 1$. Since the asymetric sigmoid contains one additional parameter, and therefore one more degree of freedom for optimization, a better fit is expected, leading to a value of $R_a^2$ higher than $R_s^2$. Therefore, the value of $\eta_1$ is expected to be lower than one. Nevertheless, a value of $\eta_1$ very close to one indicates a low increase in the squared correlation coefficient, which may not be enough to justify the increase in the model complexity.

To prevent a division by zero, if $R_s^2 = 1$ (up to $10^{-12}$ precision), then the symmetric curve is selected, as it has perfect fit and the asymmetric curve cannot yield better results. This case has been considered for robustness of the computational code; nevertheless, since this a perfect situation, it is not expected to occur in practice.

In the remaining cases, to balance between complexity and a more accurate fit, the criterion for selection follows the rule:

$$\eta_1 \leq 0.5 \ \Rightarrow \text{asymmetric curve is selected}$$
$$\qquad\qquad\qquad\qquad\qquad\qquad\qquad ( \ 29 \ )$$
$$\eta_1 > 0.5 \ \Rightarrow \text{symmetric curve is selected}$$

With this choice, a more substantial gain in the accuracy of the fit has to be obtained to justify a more complex curve.

### 2.6.2. One or multiple sigmoids

Given the number $N_s$ of sigmoids, two fits are performed, i.e., using one and using $N_s$ sigmoids. The criterion to decide upon the use of one or multiple sigmoids is similar to the one defined in the previous section for symetric and asymetric sigmoids.

We define a new ratio $\eta_2$ as

$$\eta_2 = \frac{1 - R_m^2}{1 - R_o^2} \quad (\,30\,)$$

where $R_o^2$ and $R_m^2$ represent the squared correlation coefficient obtained from fitting one and multiple sigmoids to the calibration data, respectively, provided that $R_o^2$ is different from one. As in the previous case, if $R_o^2=1$ (which is a perfect situation, not expected to occur in practice), then only one sigmoid is adopted.

The criterion to choose between one and multiple sigmoids is then:

$$\eta_2 \leq \frac{1}{N_s} \Rightarrow \text{multiple sigmoids are selected}$$

$$(\,31\,)$$

$$\eta_2 > \frac{1}{N_s} \Rightarrow \text{one sigmoid is selected}$$

### 2.7. Summary

The methodology proposed in the current paper is summarized by the flowchart presented in Figure 2.

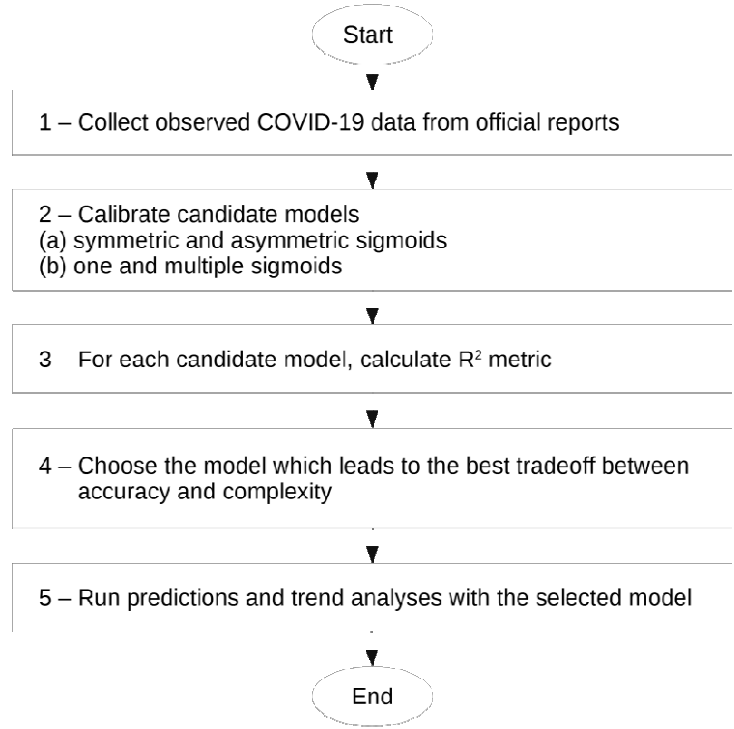

**Figure 2 – Flowchart summarizing the methodology proposed in the current paper.**

## 2.8. Implementation

The computer program described in this paper was developed using MATLAB® 2020a, with the Optimization Toolbox™ and the MATLAB Compiler™.

The program uses as data source reports published in spreadsheet format in the websites of the European Centre for Disease Prevention and Control (ECDC) [12], of Johns Hopkins University [9] and of the Brasil.io project [4].

The ECDC reports contain country-wise data of the countries in the world, while the reports of Johns Hopkins University and of the Brasil.io project presents data of United States counties and of Brazilian states and cities, respectively.

The data inform the number of newly infected and deceased people on each date. These numbers are informed separately for each region, allowing to perform an independent analysis for each of them.

## 3. Results

### 3.1. The Graphical User Interface

The computer program may be downloaded from the following link, where the data files updated until 12-Aug-2020 are also available.

**https://gitlab.com/rubensjma/sigmoid-covid-19**

The folder contains a "readme" file, which explain the main features and the preliminary steps to use the program. We emphasize that the program may be installed and run directly from the operating system, independently of the user possessing a licensed MATLAB® installation. Should the user have MATLAB® and the required packages installed, he/she may run directly a different file from the package without any installation.

A Graphical User Interface (GUI), illustrated in Figure 3 and Figure 4, will appear. These figures present the main screen of the GUI with data from the European Centre for Disease Prevention and Control and from Johns Hopkins University, respectively. A zoom was applied to these figures to allow for a better reading of their contents; that is the reason why the names of some U.S. counties appear truncated and why the predictions in the bottom of the figures appear incomplete.

When running the GUI for the first time, the user is advised to initially select the option "File: Download New Data File" of the main menu, as described in further detail below.

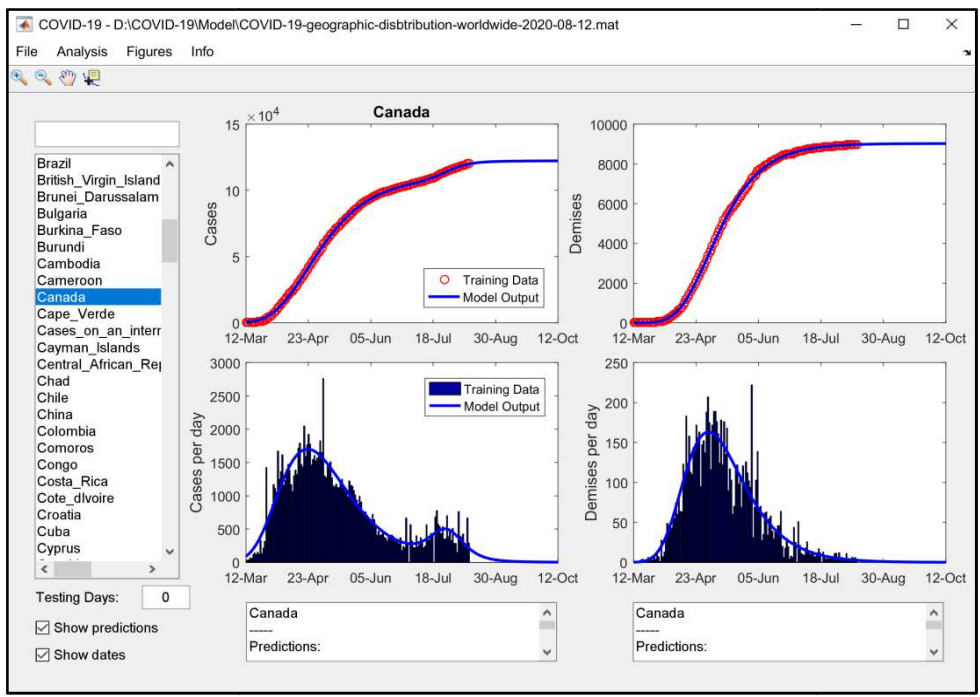

**Figure 3 – Main screen of the Graphical User Interface (GUI), with worldwide data from the European Centre for Disease Prevention and Control (ECDC). Note the list of countries on the left side.**

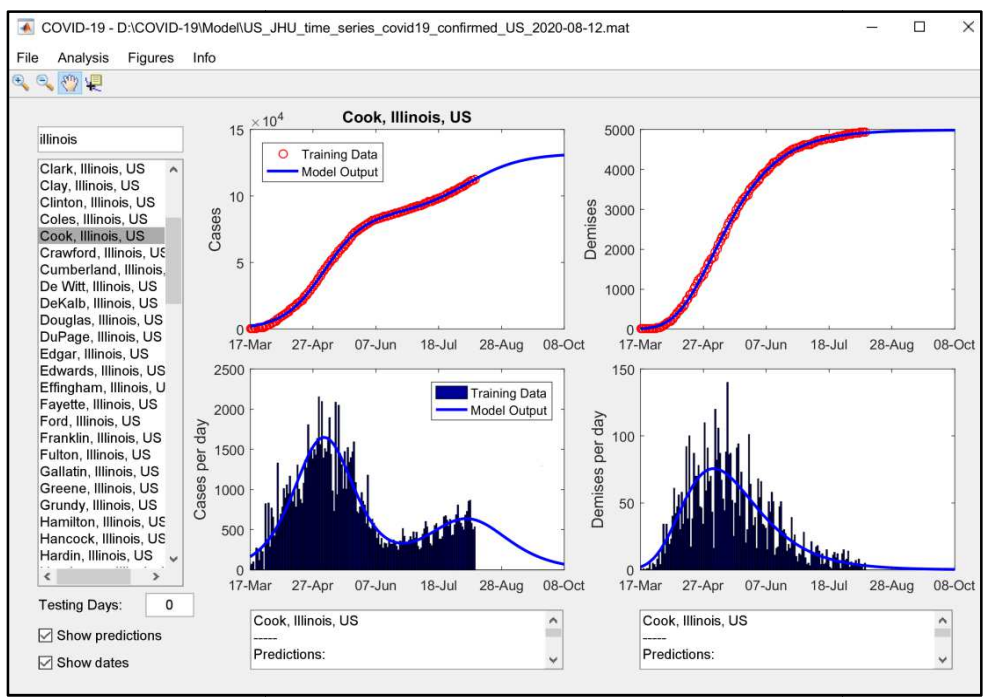

**Figure 4 – Main screen of the Graphical User Interface (GUI), with U.S. data from Johns Hopkins University. Note the list of counties on the left side and the search for "Illinois" on the upper left corner.**

The GUI contains a main menu with the following options:

- **"File: Load Data File"** – This option is used to load data from a spreadsheet in the standard formats defined by the ECDC, by Johns Hopkins University and by the Brasil.io project. Upon first reading, the spreadsheet will be converted to a MATLAB® data file with extension .mat, in order to speed up the following readings. When the interface is opened, the last data file is automatically reloaded.

- **"File: Last Data Files"** – This option is used to reload one of the last data files, as illustrated in Figure 5.

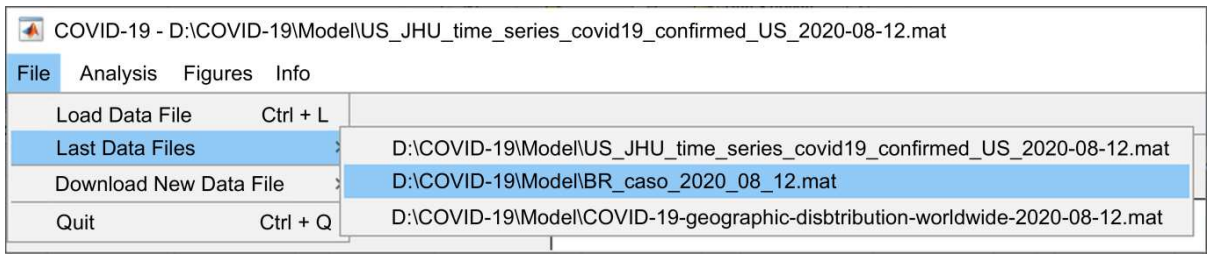

**Figure 5 – "File: Last Data Files" option from the main menu.**

- **"File: Download New Data File"** – This option shows the menu presented in Figure 6, where the user may select one of the following websites: ECDC, Johns Hopkins University or Brasil.io project. The user may choose to download the data file directly or to access one of these sites, using the default web browser. It is easier to choose the automatic download; however, we opt to also provide an option to access the websites as an acknowledgement of the work performed by the people responsible for them.

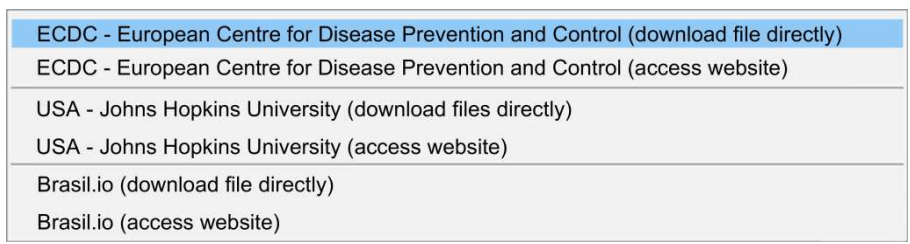

**Figure 6 – Options to download new data files and to access the corresponding websites.**

- **"File: Quit"** – This is a standard option to close the interface.

- **"Analysis: Run Trend Analysis"** – This option runs the trend analysis, as described in the previous section, and presents its results in an external figure. Examples of results of this analysis are presented in the next section.

- **"Figures: Export Figure"** – This option is used to export the graphs of the main screen to a new figure, in order to facilitate its edition and copy to external software.

- **"Figures: Close all Figures"** – This option closes all external figures.

- **"About: Info"** – This option shows updated information about the interface. It also contains an acknowledgement of the sites used as data sources.

On the left of the main screen (Figure 3 and Figure 4), there is a list of all available regions, which will be henceforth called the region list. In this list, when analyzing data from the ECDC or from Johns Hopkins University, the user can select the name of the desired country or of the desired U.S. county (in English); when analyzing data from the Brasil.io project, the user can select the name of the Brazilian states and cities (in Portuguese). Brazilian states are identified by their two-letter acronym. The names of the cities are presented without accents; for instance, the cities of "São Paulo", "Santa Bárbara d'Oeste" and "Santa Fé" are identified as "Sao Paulo", "Santa Barbara d'Oeste" and "Santa Fe", respectively.

Above the region list, there is an edit field where the user can type the name of a region to look for on the list. The user can select the region name in full or in part, and may also employ regular expressions. Furthermore, a vertical bar "|" can be used representing the "or" operator, to perform a search for more than one region; for instance: **fran|germ|italy** will restrict the countries in the list to France, Germany and Italy. An empty string is used in the search bar to restore the complete list of regions. Note the search for "Illinois" in Figure 4.

Below the region list, there are two options: "Show predictions" and "Show dates". "Show predictions" is used to enable or disable the mathematical modeling (if disabled, only historical data will be shown). If "Show dates" is disabled, then sequential numbers are shown in the graphs' axes, instead of dates.

In the lower left corner of the GUI, there is an edit field where the user can specify the number of days for testing. For instance, if the user specifies a value of 7 days, then the model is calibrated with all data available until one week before the data acquisition, and the remaining days are used to test the model, allowing a comparison between the predictions of the model and the observed data.

The main screen presented in Figure 3 and Figure 4 contains four graphs, representing the accumulated (top) and daily (bottom) number of cases (left) and demises (right). Each graph contains historical data and theoretical curves representing them. Observed data are presented using either circles (accumulated values) or bars (daily values). The model output is represented by the continuous blue line.

Below the graphs, the following predictions are presented, for either cases or demises: final number, date of maximum daily occurrences and settling date (as defined in Table 1). Furthermore, the equation of the best sigmoid (or set of sigmoids in case more than one wave is identified) matching the accumulated data is presented, as well as the indices RMSE and $R^2$, defined in the previous section. The predictions are presented in editable fields, such that the user can copy their texts and paste them in external software. An example of such information is presented in Table 2.

**Table 2 – Example of the information presented at the bottom of the main screen**

```
   Canada                                    Canada
   -----                                     -----
   Predictions:                              Predictions:
    Final number of cases: 122407             Final number of demises: 9030
    Date of maximum daily cases: 23-Apr-2020  Date of maximum daily demises: 02-May-2020
    Settling date (cases): 11-Aug-2020        Settling date (demises): 18-Jul-2020
    -----                                     -----
    Number of accumulated cases on 12-Aug-2020:  Number of accumulated demises on 12-Aug-
 120406                                       2020: 8991
    -----                                     -----
    Prediction with data from 12-Mar-2020 to  Prediction with data from 12-Mar-2020 to
 12-Aug-2020                                  12-Aug-2020
    Curve: f(t)=(109370 / (1 + 3.56e-02 exp(-  Curve: f(t)=(9030 / (1 + 5.93e-02 exp(-((t-
 ((t-42.34)/23.24)))^(1/3.56e-02)) + (13037 / (1  51.30)/19.77)))^(1/5.93e-02))
 + exp(-(t-137.34)/7.67)))                        Normalized RMSE (demises): 5.499e-03
    Normalized RMSE (cases): 4.904e-03            R^2  (demises): 0.9998
    R^2  (cases): 0.9998
```

### 3.2. Illustrative Results – Time Series

In order to illustrate the use of the tool to perform predictions, Figure 7 to Figure 9 show the model results for the country of Bolivia, the U.S. county of Los Angeles and the Brazilian capital city of Brasilia, respectively. Data updated on 12-Aug-2020 were used. The number of testing days was set to seven, meaning that the model was calibrated with data until 05-Aug-2020 and the following seven days were used to compare the predicted and observed values.

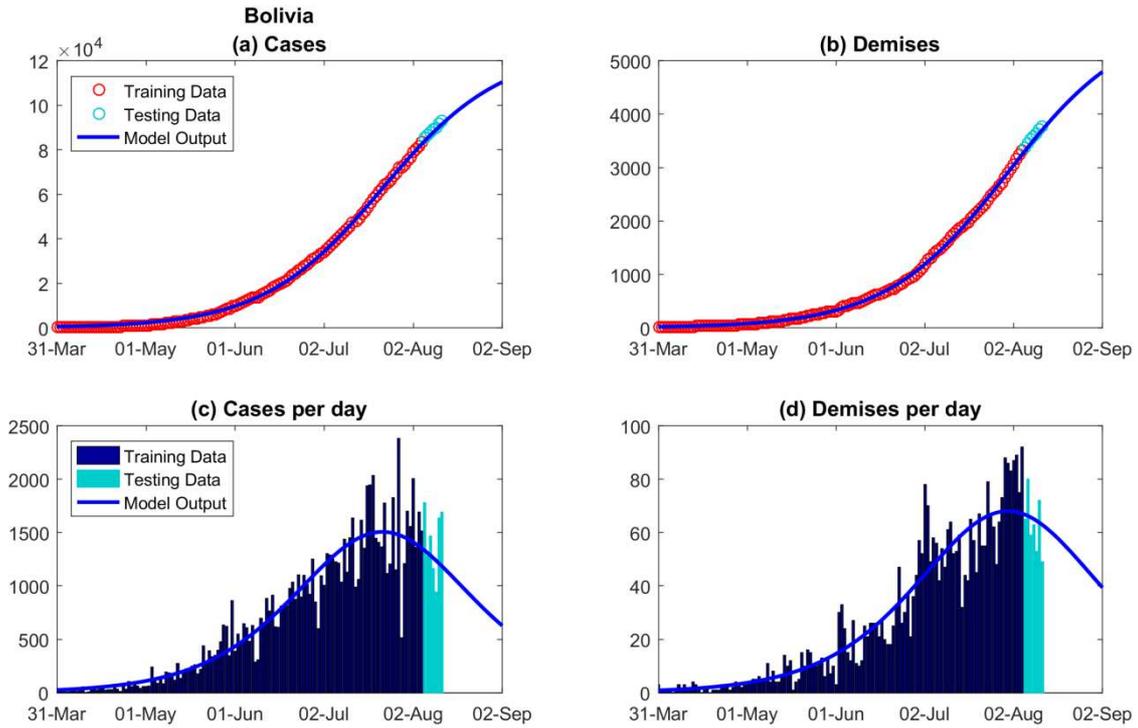

**Figure 7 – Graphs of cases and demises for the country of Bolivia. The figure shows accumulated (top) and daily (bottom) occurrences.**

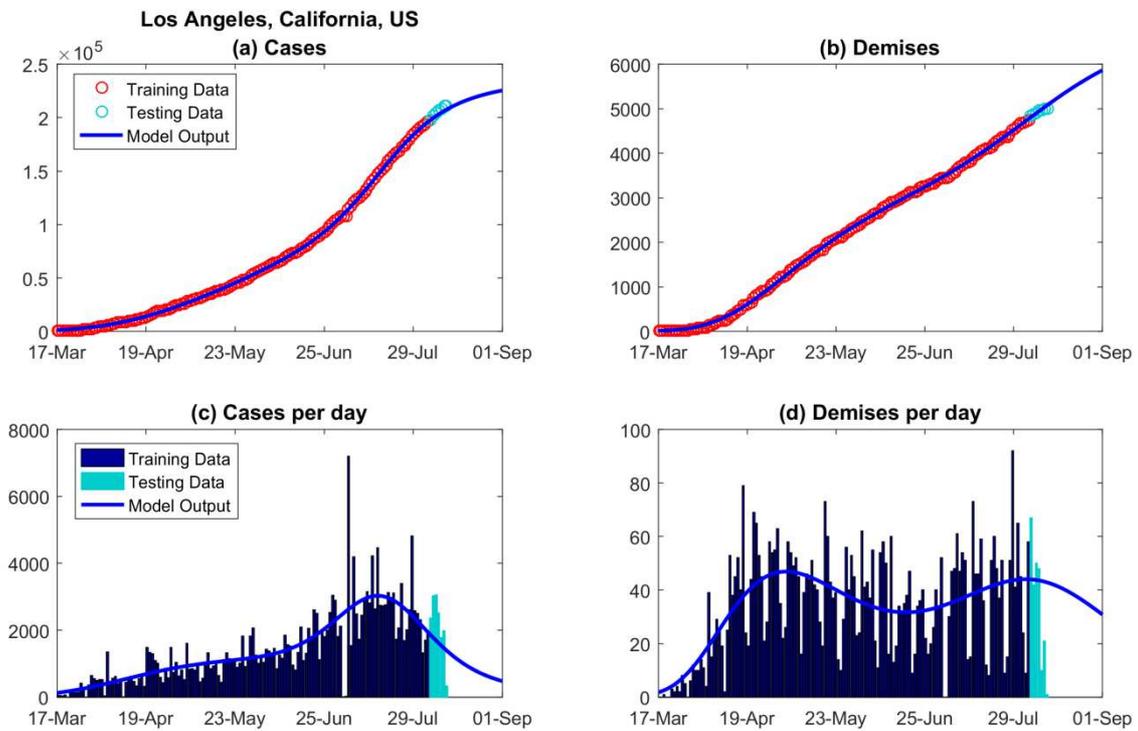

**Figure 8 – Graphs of cases and demises for the U.S. county of Los Angeles. The figure shows accumulated (top) and daily (bottom) occurrences.**

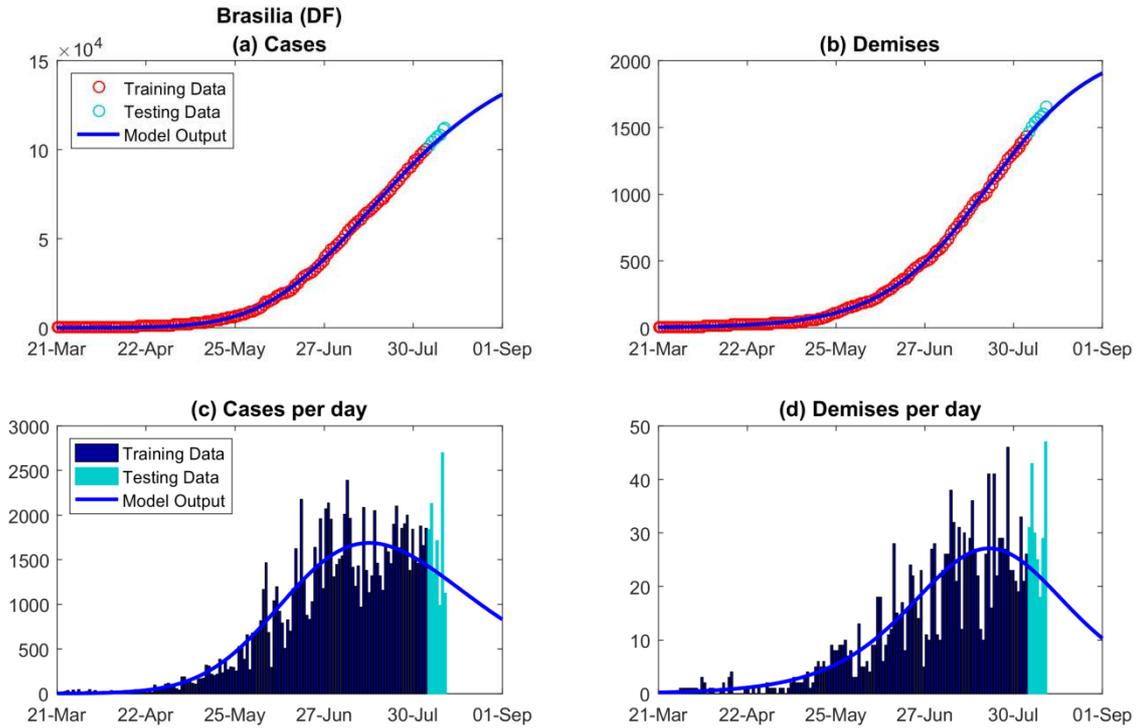

**Figure 9 – Graphs of cases and demises for the Brazilian capital city of Brasilia. The figure shows accumulated (top) and daily (bottom) occurrences.**

### 3.3. Illustrative Results – Trend Analysis

As previously mentioned, the trend analysis is run when the user selects the corresponding option in the main menu. Examples of figures resulting from such analysis are presented in Figure 10 to Figure 12, which correspond to the Brazilian city of São Paulo (SP), the US county of Cook, Illinois, and the Brazilian state of São Paulo, respectively. Each of these figures contain three subfigures, showing the predicted value of the three parameters of interest described in Table 1.

The abscissa of the graphs indicates the date of the estimation, meaning that all data available until that date were used to estimate the value of the parameter under study. It can be seen that, as expected, the values of the estimated parameters vary with the amount of data used to estimate them.

On the title of each subfigure, the values of $\gamma_7$, $\gamma_{14}$ and $\gamma_{21}$ are presented, indicating how stable each prediction is, considering the last one, two and three weeks, respectively. A value of $\gamma_k$ closer to one indicates a more stable prediction.

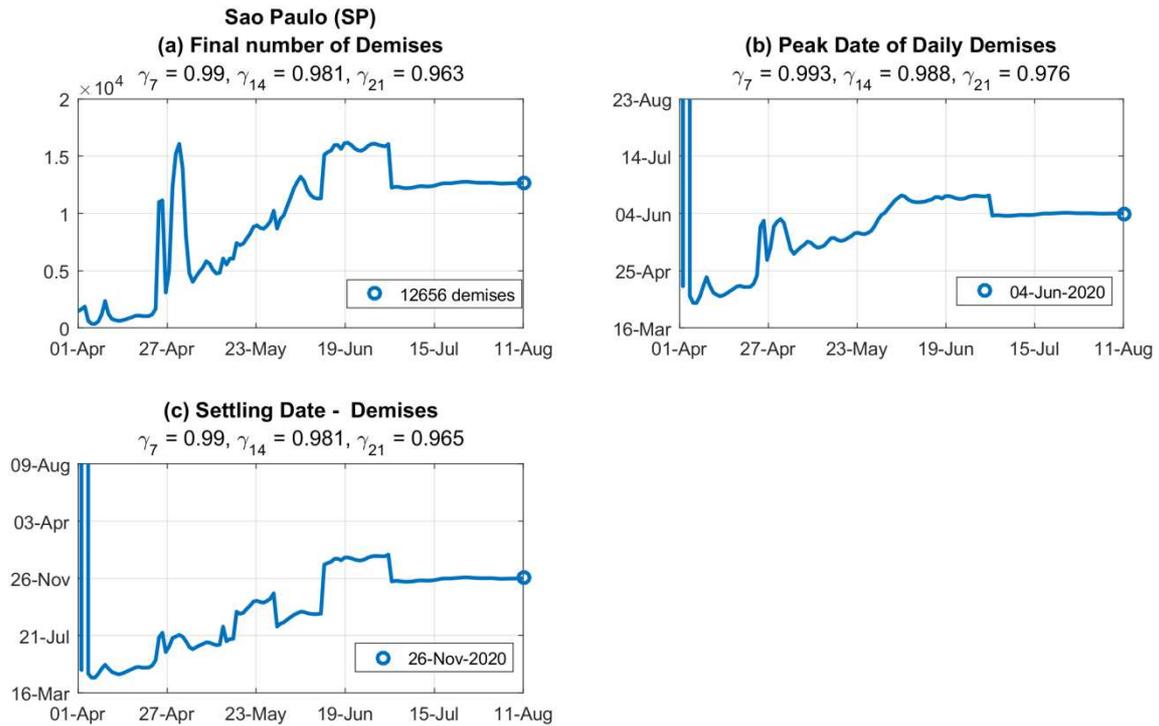

**Figure 10 – Trend analysis results for the Brazilian city of São Paulo (SP). The abscissa indicates the date of the estimation.**

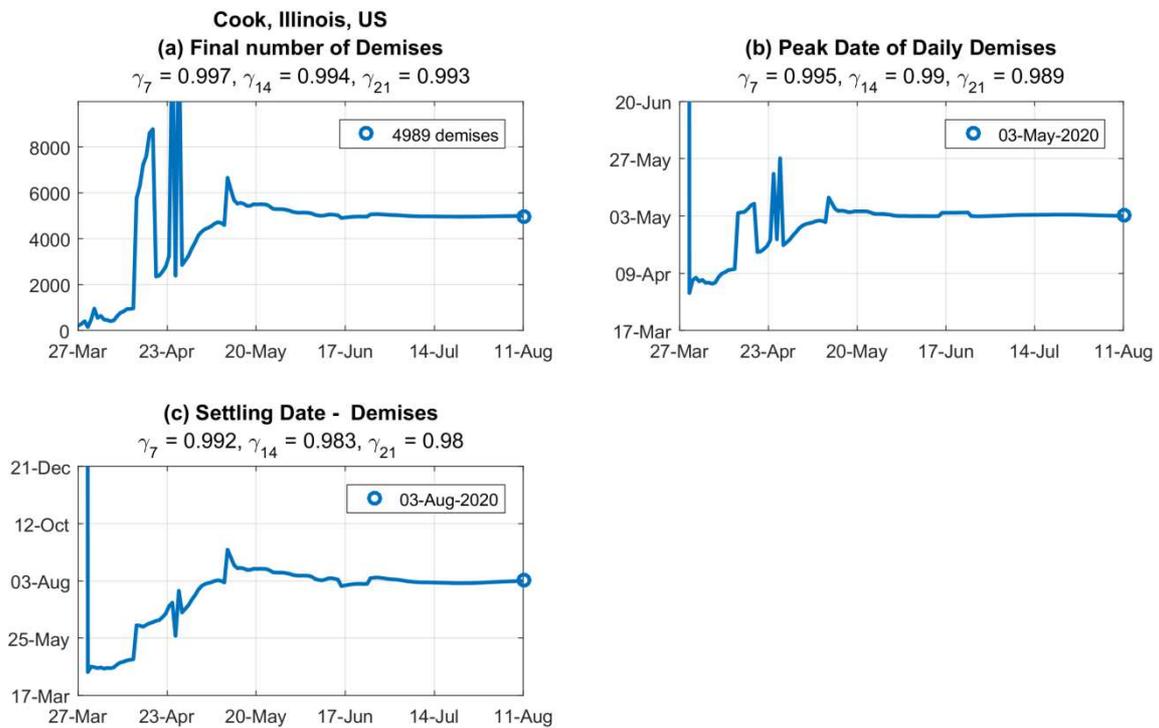

**Figure 11 – Trend analysis results for the U.S. county of Cook, Illinois. The abscissa indicates the date of the estimation.**

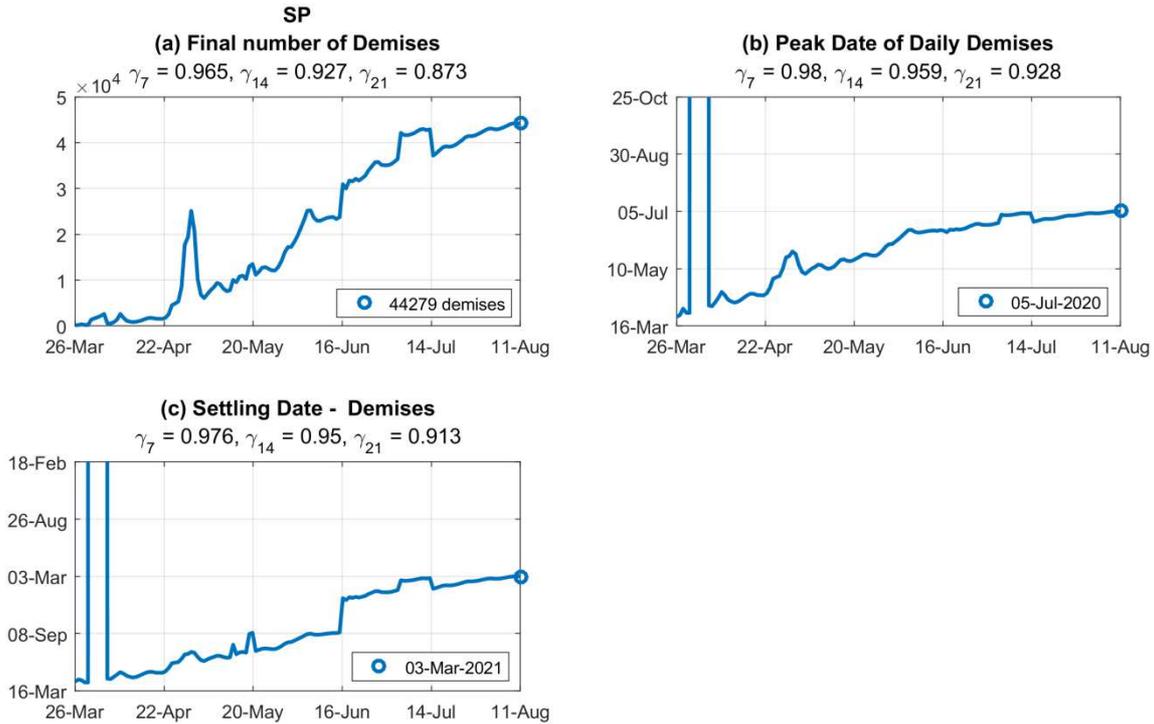

**Figure 12 – Trend analysis results for the Brazilian state of SP (São Paulo). The abscissa indicates the date of the estimation.**

## 4. Discussion

Figure 7 to Figure 9 indicate that the model represents well the training data and that the observed accumulated values (subfigures (a) and (b)) follow closely the values predicted by the model. When analyzing the daily experimental curves (subfigures (c) and (d)), it can be seen that there are high amplitude fluctuations, which may be ascribed to the nature of the observed data and may be associated to non-uniform delays in the official notifications of contaminations and deceases. These fluctuations in amplitude are similar to the sensor noise observed when analyzing physical data. In the study of dynamical systems, it is known that integral operations are robust to the presence of noise. By analogy, since our methodology uses the accumulated data (and not the daily values) to estimate the model parameters, it can be concluded that it is less sensitive to daily fluctuations in the data.

The proposed model is able to identify and represent as many epidemiological waves as necessary. For instance, the two peaks observed in the model representation of the number of daily cases in Figure 4 indicate a clear occurrence of two epidemiological waves in the U.S. county of Cook, Illinois - the daily cases were decreasing until the second week of June, and then started to consistently increase again. On the other hand, only one wave is observed in the Brazilian city of Brasilia (Figure 9(c)). To the best of the authors' knowledge, no more than two epidemiological waves have been reported in any region yet. However, it may still happen, especially if there are frequent changes in the public policies, imposing and relieving containment actions. The model is ready to represent this behaviour.

Figure 10 to Figure 12, with the results of the trend analysis, indicate how the prediction of each parameter of interest varies with time.

Figure 10-(a) represents the final number of predicted demises in the city of São Paulo (SP). It can be seen that there are oscillations in the predictions until May 5th. These oscillations result from the inclusion of new data and are expected to occur when the pandemic is at an early stage. From May 5th to June 12th, there is a clear increasing tendency in the number of demises. On June 13th, the prediction stabilizes around approximately 15000 demises. Finally, on July 2nd, the number of demises reaches approximately 12600 and is stabilized around this value ever since. Likewise, Figure 10-(b) and (c) indicate a stable prediction of the date of maximum daily demises and of the settling date since July 2nd.

Similarly, Figure 11-(a) indicates that the dynamics of the pandemic in the U.S. county of Cook, Illinois, followed a similar pattern. There were oscillations until the end of April, followed by an increasing trend until May 17th, a slightly decreasing tendency and finally an stabilized predicted value of approximately 5000 demises since June 17th.

On the other hand, Figure 12 indicates that the pandemic is not yet stabilized in the Brazilian state of SP (São Paulo). An increasing tendency can be seen over the last weeks. For instance, Figure 12-(a) shows that the predicted number of demises was approximately 37000 on July 14th and changed to 44000 on August 11th, indicating an increase of approximately 20% in 28 days.

The stability of such predictions over the last one, two and three weeks may be verified by the values of $\gamma_7$, $\gamma_{14}$ and $\gamma_{21}$ shown in the title of each figure. It can be seen that these values are close to one, indicating stabilized or near-to-stabilization predictions.

The values of $\gamma_k$ are intended to represent the convergence of the estimations. As an additional feature, they may also be used as a measurement of the quality of the prediction – higher values of $\gamma$ indicate that the pandemic has been following the same predicted behavior over the last weeks. For instance, when analysing the values of $\gamma_{21}$ presented in Figure 10-(a) and Figure 12-(a), it can be seen that such values are $\gamma_{21} = 0.963$ and $\gamma_{21} = 0.873$ for the city and for the state of São Paulo, respectively. One may infer from these numbers that the predictions obtained in the last three weeks are more stable in the city of São Paulo than in the state with the same name. This is the same conclusion that was achieved by analysing the curves, as described in the previous paragraphs.

It should be emphasized that the values in Figure 10-(a), Figure 11-(a) and Figure 12-(a) do not refer to the number of demises on the day of the analysis, but rather to the predicted final number of demises, estimated on the basis of all data available until that date. This is an innovative approach for trend analysis in this context and, to the best of the authors' knowledge, has not been proposed before.

Additionally, the same analyses presented here for the number of demises can be run for the number of infected people.

Typical trends observed in this kind of analysis are (a) oscillations, (b) increasing values and (c) stabilized values. High-amplitude oscillations may occur in the beginning of the pandemic and do not allow to reach any conclusion; however, they usually disappear after the first few weeks. Increasing values indicate a need of more compelling action by the authorities, while stabilized values indicate that the pandemic is under control.

The stabilized results for the city of São Paulo and for the county of Cook, Illinois, allow to conclude that that the actions of the local governments to control the pandemic are taking effect. It is of public interest to determine how the disease will spread in each city after the restriction measures are alleviated. For this purpose, the trend analysis should be run again. Should a new increasing tendency be observed, the authorities would be advised to reinstate some containment measures.

It is important to point out that these results, although helpful, should be validated by medical experts and not be considered alone when deciding public policies.

The trend analysis may be run for a country, a state, a county or a city. It provides more useful information when it is run for smaller administrative regions such as a county or a city, because it allows supporting decision by local authorities based on specific data of the region under consideration.

A limitation of the proposed approach is that it is not adequate to analyze the pandemic in very small cities or counties, because the number of infections and demises is usually very low, not allowing a good fitting by the mathematical model proposed here. However, for medium- and large-sized cities or counties, informative results are expected, as the ones presented here for the U.S. counties of Cook, Illinois and Los Angeles, California and for the Brazilian cities of Brasilia (DF) and São Paulo (SP).

The results presented here are illustrative and correspond to the scenario on the date when the data were acquired, that is, on 12-Aug-2020. These analyses should always employ updated data to increase their reliability. Therefore, the authors recommend these studies to be repeated periodically, at least on a weekly basis. The developed computer program allows to easily perform this task.

## 5. Conclusion

This paper proposed a methodology and a computational tool to forecast the COVID-19 pandemic throughout the world, providing useful resources for health-care authorities. A user-friendly Graphical User Interface (GUI) in MATLAB® was developed and can be downloaded online for free use. An innovative approach for trend analysis was presented.

Resources in the computational tool allow to quickly run analyses for the desired regions. Additional options allow to access the official website of the European Centre of Disease Prevention and Control, of Johns Hopkins University and of the Brasil.io project, in order to download new data as soon as they are published online. To this date, these institutions have been updating their reports on a daily basis.

The analyses run by the program are intended only as an aid and the results should be interpreted with care. They do not replace a careful analysis by experts. Nevertheless, such results may be a very useful tool to assist the authorities in their decision-making process.

The proposed program is in continuous development and future added features will be published and described in the project webpage. The authors would appreciate any feedback and suggestions to improve the computational tool.

The program, in its current version, is able to process detailed information about U.S. counties and about Brazilian states and cities. These two countries were chosen because they have continental dimensions and are currently the focus of the COVID-19 pandemic. Nevertheless, the same resource could be extended to other countries. For this purpose, the main requirement would be to write a code to read other country data files and convert them to the format recognized by the program, which is quite simple.

Future works can employ the same methodology and adapt the computer tool to describe the dynamics of other epidemics around the world. In the recent past, no pandemic was as severe as the COVID-19, but there were occurrences of other diseases such as Influenza A and MERS-CoV. Should a similar epidemic occur again, the computer program described here would be a resourceful tool.

## Acknowledgments

The authors acknowledge the European Centre for Disease Prevention and Control (ECDC), Johns Hopkins University and the Brasil.io project for making the COVID-19 data publicly available and for allowing its use for research purposes.

## References

[1]   Aleta A, Martín-Corral D, Pastore y Piontti A et al. Modelling the impact of testing, contact tracing and household quarantine on second waves of COVID-19. Nat Hum Behav (2020). https://doi.org/10.1038/s41562-020-0931-9

[2]   Atan R, Raman SA, Sawiran MS, Mohamed N, Mail R. Financial performance of Malaysian local authorities: A trend analysis. In 2010 International Conference on Science and Social Research. 2010: 271-276. https://doi.org/10.1109/cssr.2010.5773782

[3]   Barnston AG. Correspondence among the Correlation, RMSE, and Heidke Forecast Verification Measures; Refinement of the Heidke Score. Weather Forecasting. 1992. 699-709. https://doi.org/10.1175/1520-0434(1992)007<0699:CATCRA>2.0.CO;2

[4]   Brasil.io Project. http://brasil.io/. Accessed on August 12th, 2020.

[5]   Chan JF, Sridhar S, Yip CC, Lau SK, Woo PC. The role of laboratory diagnostics in emerging viral infections: the example of the Middle East respiratory syndrome epidemic. Journal of Microbiology. 2017; 55(3): 172-182. https://doi.org/10.1007/s12275-017-7026-y

[6]   Chen L, Xiong J, Bao L, Shi Y. Convalescent plasma as a potential therapy for COVID-19. Lancet Infect Dis. 2020;20(4):398-400. https://doi.org/10.1016/s1473-3099(20)30141-9

[7]   Chinazzi M, Davis JT, Ajelli M, et al. The effect of travel restrictions on the spread of the 2019 novel coronavirus (COVID-19) outbreak. Science. 2020;368(6489):395-400. . https://doi.org/10.1126/science.aba9757

[8]   Chretien JP, George D, Shaman J, Chitale RA, McKenzie FE. Influenza forecasting in human populations: a scoping review. PLoS one. 2014: 9. https://doi.org/10.1371/journal.pone.0094130

[9]   Dong E, Du H, Gardner L. An interactive web-based dashboard to track COVID-19 in real time. The Lancet infectious diseases, 2020: 20(5), 533-534. https://doi.org/10.1016/s1473-3099(20)30120-1

[10]  Dorf RC, Bishop RH. Modern Control Systems. 13th ed. London: Pearson; 2016.

[11] Dugas, A. F., Jalalpour, M., Gel, Y., Levin, S., Torcaso, F., Igusa, T., & Rothman, R. E. Influenza forecasting with Google flu trends. PloS one. 2013: 8. https://doi.org/10.1371/journal.pone.0056176

[12] European Centre for Disease Prevention and Control (ECDC). Download today's data on the geographic distribution of COVID-19 cases worldwide. https://www.ecdc.europa.eu/en/publications-data/download-todays-data-geographic-distribution-covid-19-cases-worldwide. Accessed on August 12th, 2020.

[13] Geng HY, Tan, WJ. A novel human coronavirus: Middle East Respiratory Syndrome human Coronavirus. Science China Life sciences. 2013; 56(8): 683-687. https://doi.org/10.1007/s11427-013-4519-8

[14] Gill PE, Wong E. Sequential Quadratic Programming Methods. In: Lee J, Leyffer S, editors. Mixed Integer Nonlinear Programming. New York: Springer; 2012. pp. 147–224. https://doi.org/10.1007/978-1-4614-1927-3_6

[15] Guan WJ, Ni ZY, Hu Y, et al. Clinical Characteristics of Coronavirus Disease 2019 in China. N Engl J Med. 2020;382(18):1708-20. https://doi.org/10.1056/nejmoa2002032

[16] Guastalegname M, Vallone A. Could chloroquine /hydroxychloroquine be harmful in Coronavirus Disease 2019 (COVID-19) treatment? [published online ahead of print, 2020 Mar 24]. Clin Infect Dis. 2020;321. https://doi.org/10.1093/cid/ciaa321

[17] Hernandez-Matamoros A, Fujita H, Hayashi T, Perez-Meana H. Forecasting of COVID19 per regions using ARIMA models and polynomial functions. Applied Soft Computing, 2020, 106610. https://doi.org/10.1016/j.asoc.2020.106610

[18] Hou C, Chen J, Zhou Y, Hua L, Yuan J, He S, Zhang J. The effectiveness of quarantine of Wuhan city against the Corona Virus Disease 2019 (COVID-19): A well-mixed SEIR model analysis. Journal of medical virology, 2020. https://doi.org/10.1002/jmv.25827

[19] Huang C, Wang Y, Li X, et al. Clinical features of patients infected with 2019 novel coronavirus in Wuhan, China [published correction appears in Lancet. 2020 Jan 30]. Lancet. 2020;395(10223):497-506. https://doi.org/10.1016/S0140-6736(20)30183-5

[20] Khan WU, Ye Z, Chaudhary NI, Raja MAZ. Backtracking search integrated with sequential quadratic programming for nonlinear active noise control systems. Applied Soft Computing, 2018, 73, 666-683. https://doi.org/10.1016/j.asoc.2018.08.027

[21] Khalilpourazari S., Pasandideh SHR, Niaki STA. Optimization of multi-product economic production quantity model with partial backordering and physical constraints: SQP, SFS, SA, and WCA. Applied Soft Computing, 2016, 49, 770-791. https://doi.org/10.1016/j.asoc.2016.08.054

[22] Kim Y, Lee S, Chu C, Choe S, Hong S, Shin, Y. The characteristics of Middle Eastern respiratory syndrome coronavirus transmission dynamics in South Korea. Osong public health and research perspectives. 2016; 7(1), 49-55. https://doi.org/10.1016/j.phrp.2016.01.001

[23] Lagarias JC, Reeds JA, Wright MH, Wright PE. Convergence properties of the Nelder--Mead simplex method in low dimensions. SIAM Journal on optimization. 1998, 9(1), 112-147. https://doi.org/10.1137/S1052623496303470

[24] Leung K, Wu JT, Liu D, Leung GM. First-wave COVID-19 transmissibility and severity in China outside Hubei after control measures, and second-wave scenario planning: a modelling impact assessment. The Lancet. 2020; 395(10223): 1382-1393. https://doi.org/10.1016/s0140-6736(20)30746-7

[25] Lin Q, Zhao S, Gao D, Lou Y, Yang S, Musa SS, Wang M, Cai Y, Wang W, Yang L, He D. A conceptual model for the coronavirus disease 2019 (COVID-19) outbreak in Wuhan, China with individual reaction and governmental action. International journal of infectious diseases. 2020a; 93: 211-216. https://doi.org/10.1016/j.ijid.2020.02.058

[26] Lin YF, Duan Q, Zhou Y, Yuan T, Li P, Fitzpatrick T et al. Spread and impact of COVID-19 in China: a systematic review and synthesis of predictions from transmission-dynamic models. Frontiers in medicine, 2020b; 7, 321. https://doi.org/10.3389/fmed.2020.00321

[27] Linton NM, Kobayashi T, Yang Y, et al. Incubation Period and Other Epidemiological Characteristics of 2019 Novel Coronavirus Infections with Right Truncation: A Statistical Analysis of Publicly Available Case Data. J Clin Med. 2020;9(2):538. https://doi.org/10.1101/2020.01.26.20018754

[28] Longobardi A, Villani, P. Trend analysis of annual and seasonal rainfall time series in the Mediterranean area. International journal of Climatology. 2010; 30(10): 1538-1546. https://doi.org/10.1002/joc.2001

[29] López L, Rodó X. The end of social confinement and COVID-19 re-emergence risk. Nat Hum Behav 4, 746–755 (2020). https://doi.org/10.1038/s41562-020-0908-8

[30] Lundstrom K. Coronavirus Pandemic - Therapy and Vaccines. Biomedicines 2020, 8, 109. https://doi.org/10.3390/biomedicines8050109

[31] Mathworks. Documentation of the fmincon function. Available at https://www.mathworks.com/help/optim/ug/fmincon.html. 2020. Acessed on August 12th, 2020.

[32] Mohamadou Y, Halidou A, Kapen, PT. A review of mathematical modeling, artificial intelligence and datasets used in the study, prediction and management of COVID-19. Applied Intelligence, 2020, 1-13. https://doi.org/10.1007/s10489-020-01770-9

[33] Mousavizadeh A, Dastoorpoor M, Naimi E, Dohrabpour K. Time-trend analysis and developing a forecasting model for the prevalence of multiple sclerosis in Kohgiluyeh and Boyer-Ahmad Province, southwest of Iran. Public health, 2018; 154, 14-23.

[34] Nishiura H. Real-time forecasting of an epidemic using a discrete time stochastic model: a case study of pandemic influenza (H1N1-2009). BioMed EngOnLine. 2011; 10: 15. https://doi.org/10.1186/1475-925x-10-15

[35] Nocedal J, Wright SJ. Numerical Optimization. 2nd ed. New York: Springer; 2006. https://doi.org/10.1007/b98874

[36] Oliveira PT, Santos e Silva CM, and Lima KC. Climatology and trend analysis of extreme precipitation in subregions of Northeast Brazil. Theoretical and Applied Climatology 130.1-2 (2017): 77-90. https://doi.org/10.1007/s00704-016-1865-z

[37] Ozma MA, Maroufi P, Khodadadi E, et al. Clinical manifestation, diagnosis, prevention and control of SARS-CoV-2 (COVID-19) during the outbreak period. Infez Med. 2020 Ahead of print Jun 1;28(2):153-165.

[38] Paiva HM, Afonso RJM, de Oliveira IL, and Garcia GF. A data-driven model to describe and forecast the dynamics of COVID-19 transmission. PloS one. 2020; 15(7), e0236386. https://doi.org/10.1371/journal.pone.0236386

[39] Phelan AL. COVID-19 immunity passports and vaccination certificates: scientific, equitable, and legal challenges [published online ahead of print, 2020 May 4]. Lancet. 2020;S0140-6736(20)31034-5. https://doi.org/10.1016/s0140-6736(20)31034-5

[40] Rath S, Tripathy A, Tripathy AR. Prediction of new active cases of coronavirus disease (COVID-19) pandemic using multiple linear regression model. Diabetes & Metabolic Syndrome: Clinical Research & Reviews. 2020. https://doi.org/10.1016/j.dsx.2020.07.045

[41] Ribeiro MHDM, da Silva, RG, Mariani VC, & dos Santos Coelho, L. (2020). Short-term forecasting COVID-19 cumulative confirmed cases: Perspectives for Brazil. Chaos, Solitons & Fractals. 2020, 109853. https://doi.org/10.1016/j.chaos.2020.109853

[42] Richards FJ. A flexible growth function for empirical use. Journal of experimental Botany. 1959, 10(2), 290-301. https://doi.org/10.1093/jxb/10.2.290

[43] Ross R. CDC estimate of global H1N1 pandemic deaths: 284,000. Center for Infectious Disease Research and Policy. 2012. https://www.cidrap.umn.edu/news-perspective/2012/06/cdc-estimate-global-h1n1-pandemic-deaths-284000. Accessed on May 11th 2020.

[44] Salgotra R, Gandomi M, Gandomi AH. Time Series Analysis and Forecast of the COVID-19 Pandemic in India using Genetic Programming. Chaos, Solitons & Fractals, 2020; 109945. https://doi.org/10.1016/j.chaos.2020.109945

[45] Sarkar M, Agrawal AS, Dey RS, Chattopadhyay S, Mullick R, De P, Chakrabarti S, Chawla-Sarkar M. Molecular characterization and comparative analysis of pandemic H1N1/2009 strains with co-circulating seasonal H1N1/2009 strains from eastern India. Archives of virology. 2011; 156(2): 207-217. https://doi.org/10.1007/s00705-010-0842-6

[46] Soares SCM, dos Santos KMR, de Morais Fernandes FCG, Barbosa IR, de Souza DLB. Testicular Cancer mortality in Brazil: trends and predictions until 2030. BMC Urology, 2019, 19(1), 59. https://doi.org/10.1186/s12894-019-0487-z

[47] Steyerberg EW. Overfitting and optimism in prediction models. Clinical Prediction Models. Springer, Cham. 2019. 95-112. https://doi.org/10.1007/978-0-387-77244-8_5

[48] Wen M, Li P, Zhang L, Chen Y. Stock Market Trend Prediction Using High-Order Information of Time Series. IEEE Access, 2019; 7, 28299-28308. https://doi.org/10.1109/access.2019.2901842

[49] World Health Organization (WHO). Coronavirus disease (covid-19). Situation Report – 51 http://www.who.int/docs/default-source/coronaviruse/situation-reports. Acessed 09 May 2020.

[50] World Health Organization (WHO). Coronavirus disease (covid-19). Situation Report – 110 http://www.who.int/docs/default-source/coronaviruse/situation-reports/.Acessed 09 May 2020.

[51] World Health Organization (WHO). Novel coronavirus – China. http://www.who.int/csr/don/12-January-2020-novel-coronavirus-china/en/.Acessed 09 May 2020.

[52] Wu JT, Leung K, Leung GM. Nowcasting and forecasting the potential domestic and international spread of the 2019-nCoV outbreak originating in Wuhan, China: a modelling study [published

correction appears in Lancet. 2020 Feb 4]. Lancet. 2020; 395(10225):689-97. https://doi.org/10.1097/01.ogx.0000688032.41075.a8

[53] Xu S, Li Y. Beware of the second wave of COVID-19. Lancet. 2020; 395(10233), 1321–1322. https://doi.org/10.1016/S0140-6736(20)30845-X

[54] Yang, Zifeng, et al. Modified SEIR and AI prediction of the epidemics trend of COVID-19 in China under public health interventions. Journal of Thoracic Disease, 2020; 12 (3), 165. https://doi.org/10.21037/jtd.2020.02.64

[55] Yuan H, Li X, Wan G, Sun L, Zhu X, Che F, Yang, Z. Type 2 diabetes epidemic in East Asia: a 35–year systematic trend analysis. Oncotarget, 2018; 9(6), 6718. https://doi.org/10.18632/oncotarget.22961

[56] Zahmatkesh B, Keramat A, Alavi N, Khosravi A, Kousha A, Motlagh AG, Chaman R. Breast cancer trend in Iran from 2000 to 2009 and prediction till 2020 using a trend analysis method. Asian Pacific Journal of Cancer Prevention, 2016; 17(3), 1493-1498. https://doi.org/10.7314/apjcp.2016.17.3.1493

[57] Zhao J, Zuo T, Zheng R, Zhang S, Zeng H, Xia C., Chen W. Epidemiology and trend analysis on malignant mesothelioma in China. Chinese Journal of Cancer Research, 2017, 29(4), 361. https://doi.org/10.21147/j.issn.1000-9604.2017.04.09